\documentclass[aps,twocolumn,showkeys]{revtex4}

\pdfoutput=1
                                
\usepackage{graphicx}                                       
\usepackage{amssymb}

\newcommand{\umafigura}[3]{   
  \begin{figure}[!t]
    \centering
    \includegraphics[width=2.5in]{#1}
    \caption{#3}
    \label{#2}
  \end{figure}
}

\savebox{1}[4in]{\noindent\hskip 1.1in(a)\hskip 1.4in(b)\hfill\ }

\newcommand{\eqt}[2]{
    \begin{equation}
             {#2}
             \label{#1}
    \end{equation}
}

\newcommand{\eqp}[5]{
   \setlength{\arraycolsep}{0.0em} 
   \begin{eqnarray}
       \label{#1}
       \ds{#2}&{}={}&\ds{#3}\nonumber\\ 
        &&{#4}\:\ds{#5}
   \end{eqnarray}
}

\def\ds{\displaystyle}

\begin{document}

\title{Power-law Interpolation of AC-DC Differences}


\author{
  F. A. Silveira${}^{}$\footnote{Email address: {\it fsilveira@inmetro.gov.br} (corresponding author)},
  Regiane M. Souza${}^{}$\footnote{Email address: {\it rmsouza@inmetro.gov.br}} and
  R\'egis P. Landim${}^{}$\footnote{Email address: {\it rplandim@inmetro.gov.br}}
}

\affiliation{
  Instituto Nacional de Metrologia, Qualidade e Tecnologia,\\
  Avenida N. S. das Gra\c cas 50, 25250-020 D. Caxias RJ, Brazil}

\date{\today}


\begin{abstract}
    In this work we present a general method, commonly applied to the numerical 
    analysis of stochastic models, to interpolate AC-DC differences (usually denoted 
    by the greek letter $\delta$) between calibration points in thermal transfer standards.
    This method assigns a power-law behaviour to AC-DC differences, solely under the
    assumption that $\delta$ must be some smooth-varying function of voltage and 
    frequency. We argue it may 
    be straightfowardly applied to all working ranges of 
    the standards, with no distinction.
\end{abstract}


\maketitle

\section{Introduction}
\label{intro}

    Recent progress has been made towards the standardization of the AC voltage 
    and current \cite{laiz2005}, but thermal transference methods, 
    by which AC-volt and AC-amp\`ere are compared to standard DC functions by
    thermal conversion, remain the most important resource in AC metrology practice on
    industrial and scientific electrical standards. The measurand is the 
    difference between a DC current and the AC current needed to dissipate the 
    same quantity $Q$ of heat on the transfer standard; it is usually called 
    {\it AC-DC difference}, and denoted by the greek letter $\delta$
    \cite{inglis1992,calibration}.

    We expect the measured AC-DC differences to be a continuous function of $Q$.
    Both numerical and theoretical results, and simple considerations on the
    physics of the heat production on the converters support this assumption. Neverthless, 
    the same physical considerations are not yet able to make satisfactory 
    predictions to a metrological level.

    Despite this, we show in this work we can make use of the 
    smoothness of $\delta$ (see Sec. \ref{method}), to associate correction 
    exponents to the solution of the limiting equation for the problem, which is 
    very easily formulated, and obtain a precise form for $\delta$, at least on some 
    neighbouring frequencies and values of the input function
    \cite{cpem}.

    Sections \ref{TC} and \ref{characteristics} show results regarding the 
    AC-DC difference in thermal converters, and review some important equations that 
    relate to these    results. Sections \ref{method} and \ref{pmjtc} present a 
    detailed description  of the method, and some results obtained on transfer standards 
    and thin-film multijunction systems; a brief conclusion is then presented in the 
    last section.

    \section{Thermal Converters}
    \label{TC}

     The most simple available thermal converter is the
    {\it single junction thermal converter} (SJTC), extensively studied and
    introduced as a transfer standard in the 50's by Hermach \cite{hermach1952}.
    The SJTC may be combined with a resistor in order to extend its transfer ranges.
    This set makes a thermal converter (TC), which can be in series or parallel 
    configuration, depending on the function being calibrated, whether it is 
    voltage (when the set is assigned the term thermal voltage converter, 
    or TVC) or current (alternatively, thermal current converter, or TCC), respectively.

    A SJTC consists  of a thermocouple welded to the middle of a low-resistence
    heater wire, through which electrical current flows and dissipate heat.The weld between
    thermocouple and heater is made from ceramic or similar material, and
    provides electrical, but no thermal insulation, allowing the hot junction to
    quickly get into thermal equilibrium with the center of the heater \cite{inglis1992}. 
    These thermoelements, as they are also called, are usually mounted in an evacuated 
    glass bulb, as shown in Fig. \ref{SJTC},
    and serve well to illustrate the principles of AC-DC metrology through thermal converters.

        \umafigura{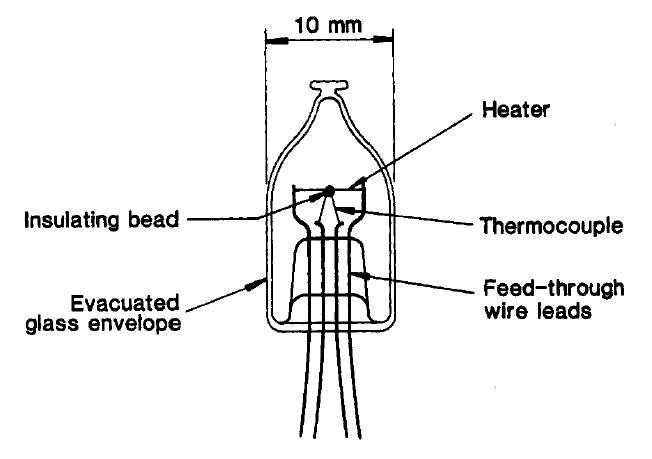}{SJTC}{{\it SJTC schematics.} The temperature
        in the heater during operation is about $220 ^oC$, and its output voltage can range
        tipically from 8 to 12 mV; the bead
        provides electrical, but no thermal insulation, and allows the hot junction to
        quickly get into thermal equilibrium with the center of the heater. After
        Ref. \cite{inglis1992}.} 
        \umafigura{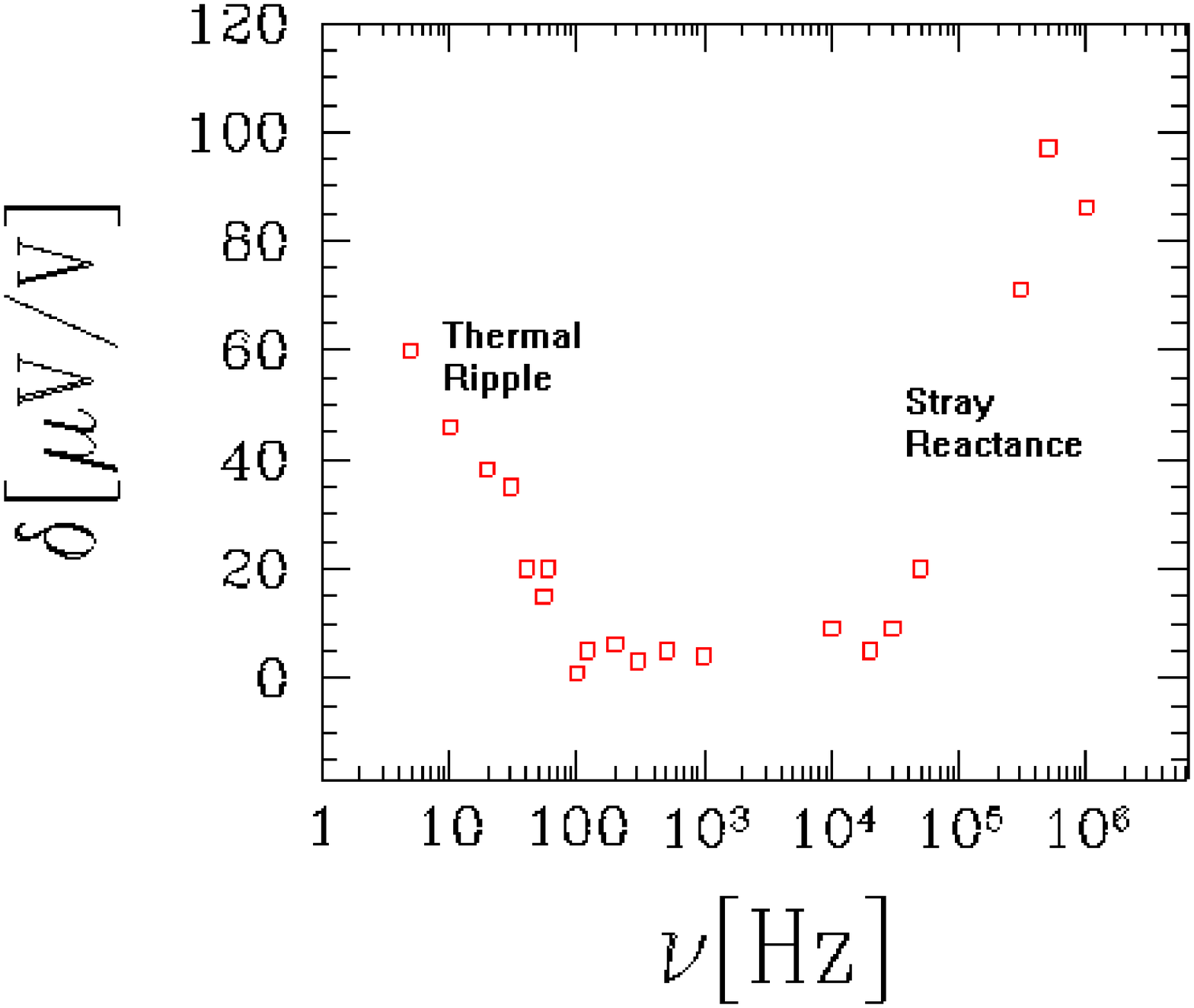}{DELTACHAR}{{\it TVC/TCC characteristics sketch.}
        This illustration shows a typical frequency behaviour of voltage AC-DC
        differences at fixed voltage, obtained from a rough simulation. The 
        main causes of AC-DC difference in SJTC are DC offset, stray reactance 
        and low-frequency thermal ripple.}

    Fig. \ref{DELTACHAR} shows  sketch of the typical frequency dependence
    of a SJTC, in which we can see the most relevant sources of error in
    thermal converters in general.
    The main causes of AC-DC difference in thermal converters are DC offset, stray 
    reactance and low-frequency thermal ripple. The 
    contribution of frequency combined effects to 
    the AC-DC difference can be of the order of 0.1 $\mu$V/V from 100 Hz to 20 kHz, to 100 $\mu$V/V at
    1 MHz for an ordinary SJTC \cite{sasaki1999}.

    Thermal conversion is today the most economically feasible form available 
    for making AC-DC transfers. Another very commonly used type of TC are 
    eletronic, solid-state-based thermal standards as the one sketched in Fig. \ref{TRUERMS}.
    The output of such an electronic TC
    responds linearly to temperature increases in a heater resistor. The output
    voltage at full input is about 2 V, instead of the 12 mV obtained from
    thermocouple-based TC's \cite{calibration,apnote}.

    Three electronic thermal standards Fluke model 
    792A\footnote{Certain commercial equipment, instruments, or materials are 
    identified in this report to facilitate understanding. 
    Such identification 
    does not imply recommendation or endorsement by Inmetro, 
    nor does it imply that the materials or equipment that are identified are 
    necessarily the best available for the purpose.} (F792A), 
    one of them assigned primary, are the transfer standards at Inmetro.
    And though studies are being carried out in order to replace them by
    thin-film multijunction converters (PMJTC) in primary standardization \cite{renato2009}, 
    such electronic transfer standards are widely used  in all areas where AC 
    transfer is required, including many other
    National Metrology Institutes (NMI).

        \umafigura{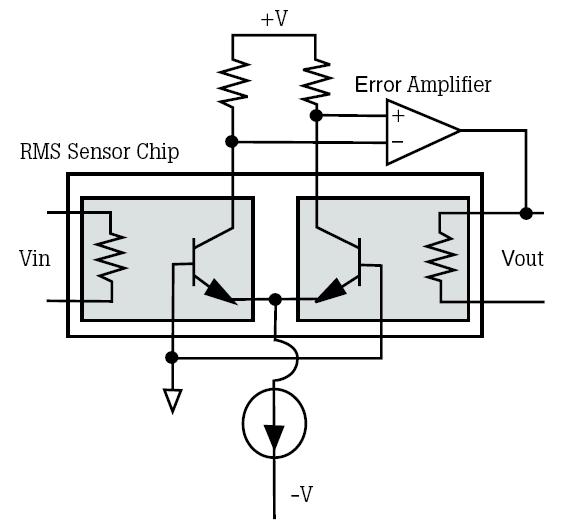}{TRUERMS}{{\it Simplified Fluke rms sensor.}
        The sensor chip consists of two matched thermal voltage converters
        that lay on the temperature sensitivity of solid-state junctions
        rather than the ordinary thermocouple. After Ref. \cite{apnote}.}


\section{Transfer Characteristics}
\label{characteristics}

    We can assign the output of a thermal converter the general form
    \eqt{TCCHAR}{
                           E=k f[Y;\omega],
    }
    \noindent where $E$ is the rms output voltage, $k$ is some constant scale factor 
    dependent on the input funtion (voltage or current)
    and $f$ is a response functional dependent on the input function time evolution $Y$ and,
    by an obvious extension, the frequency $\omega$. The functional $f$ (which may be also
    called {\it transfer characteristic}) gives, without
    loss of generality, a measure of the heat dissipated in the heater of the TC,
    whatever design it might have, be it a simple resistive element or some complex
    electronic circuit.

    May the input function $Y=I(t)$ be current, the AC-DC difference of a converter 
    at some $I$ and $\omega$, referenced to a known DC current $I_0$, is given by
    \cite{schoe1978}
    \eqt{OLDHAMDELTA}{
           \delta=-\left.{E[I]-E[I_0]\over n\cdot E[I_0]}\right|_{I=I_0}
    }
    \noindent where $n$ is the characteristic exponent (often called 
    ``normalized index" or ``linearity coefficient") of the converter.

    Should Joule effect be the only significant contribution to heat 
    generation in the converter, supposing perfect sinewave voltage input $Y=I\sin(\omega t)$,
    we should have
    \eqt{FUNCTIONALSJTC}{
          f[Y]=\int_{\omega/2\pi} dt'\cdot Y^2(t'),
    }  
    and we expect $\ds f[Y]\sim I^2$. 
    As other thermoelectric effects become important, we must consider higher order 
    contributions to $f$ in its arguments, or even incorporate time-dependent factors, if good 
    estimates for AC-DC differences are to be obtained.


    Then, following Schoenwetter \cite{schoe1978}, the lack of 
    square-law response in SJTC for sinewave input currents at low-frequencies 
    can be modelled as
    \eqp{SCHOE_INPUT}{Y^2}
            {aI^2\left(1-m\cos(2\omega t)\right)}
            {+}{ab I^4\left(1-m\cos(2\omega t)\right)^2}
    \noindent where $a,b$ are scale constants, $I$ is the rms value of  the current 
    through the heater, $\tau$ is the thermal integration time constant of the TC and
    $$
     m^2={1\over 1+ \left(2\omega\tau\right)^2    }.
    $$ 
    Substituting Eq. \ref{SCHOE_INPUT} 
    in Eq. \ref{FUNCTIONALSJTC} and carrying on the integration:
    
    \eqt{OLDHAMCHAR}{
                 f[I]= I^2 \left[\left({m^2\over 2}+1\right)bI^2+1\right]
    }
    \noindent if $\omega\neq 0$. At DC regimes ($\omega=0$), the same procedure leads to
    \eqt{OLDHAMCHAR2}{
                 f[I_0]= I^2 \left[b I^2+1\right].
    }

    According to Ref. \cite{schoe1978}, $b$ can be evaluated through DC 
    measurements of the output voltage of the TC.
    Then, applying Eqs. \ref{OLDHAMCHAR} and \ref{OLDHAMCHAR2} to Eq. \ref{OLDHAMDELTA},
    the low-frequency AC-DC difference of 
    a converter at some  $I$ and $\omega$, referenced to  a known DC standard $I_0$, 
    is given by
    \eqt{OLDHAMDELTA2}{
       \delta\approx -{bm^2\over 4}I^2.
    }

    Eqs.  \ref{SCHOE_INPUT} to \ref{OLDHAMCHAR2} also enclose the following idea: 
    losses and temperature-dependent coefficients in thermal and electrical conductivity
    of the heater increase with increasing input power, and affect the dissipated power
    in uneven ways at input signal maxima and minima. Eqs. \ref{OLDHAMCHAR} and \ref{OLDHAMCHAR2} 
    show that at low-frequency AC regimes, these non-linearities affect the 
    average power generated in  the sensor, in the sense that more AC than DC is 
    required in order to obtain the same output response.

    Eq.  \ref{OLDHAMDELTA2} was obtained by considering, as an aproximation, 
    some time-independent transfer characteristic $f$, and a frequency-tracking
    correction to the input function of the form $1+m\cos(2\omega t)$, plus
    higher order terms that account for the non-linearities discussed above. The same
    aproximation also applies to F792A measurements \cite{calibration,apnote},
    and is very similar, as far as the powers of $\nu$ and $V$ are
    concerned, to the results Hermach had earlier obtained
    \cite{hermach1952}, which account for radiation losses contributions as a perturbation 
    to the heat-conduction equation above the thermal integration time-constant limit.
    Comments on the agreement of both results to PMJTC data can be found in Ref. 
    \cite{laiz1999}.

    These results bring, among many other implications, better knowledge of
    the expected limiting curves of $\delta$ at fixed voltage and frequency at low frequencies.
    This allows us to derive formulae for interpolation of AC-DC differences between 
    calibration points.

    Measurement of AC-DC differences in voltages and frequencies different
    from the ones usually offered in calibration services
    can be an important matter for data analysis and prediction, for instance. This 
    brings the question as to whether it is
    possible to improve the results for standardization between calibration points
    of the standard, and has provided the initial motivation to this work.

\section{Power-Law Interpolation}
\label{method}

   Based on the behaviour of the AC-DC difference $\delta$ reported in the previous sections,
   we expect, as $\nu\to 0$, that

   \eqt{LIMITDELTA}{
                        \delta \sim {Y^2\over \nu^2}
   }
   but for a multiplicative constant.

   Assuming Eq. \ref{LIMITDELTA} holds, a common
   way to interpolate $(\nu,\delta)$ or  $(Y,\delta)$ pairs to
   experimental AC-DC difference data is to operate a change of variables so as to make 
   $\delta$ {\it linear} in its arguments; in few words, this means we pick $Y^2$ 
   and $1/\nu^2$ as parameters. 
   This change of variables is reported  in Ref. \cite{792apendixc}.

   Such linearization processes are
   based on some truncated integer-power expansion of $f$ in its arguments, true
   only in the limit. Deviations of Eq. \ref{LIMITDELTA} can be particularly important 
   where uncertainties
   are of the order of the interpolation error estimates. And while this is 
   not always true for the F792A, it certainly applies to 
   PMJTC standards.

   Our proposal is to neglect complex details as the limiting form of $f$
   or the random effects of fluctuations,
   while still considering the possible strong higher order terms
   mentioned earlier, by assuming, in some neighborhood of the 
   point to be interpolated, low-frequency $\delta$ has the power-law
   form

   \eqt{SCALINGDELTA}{	
                           \delta\sim {Y^{2-\beta}\over \nu^{2-\alpha}}
   }

   \noindent where $\alpha$ and $\beta$ are the corrections to the limiting
   form in Eq. \ref{LIMITDELTA}.

      \umafigura{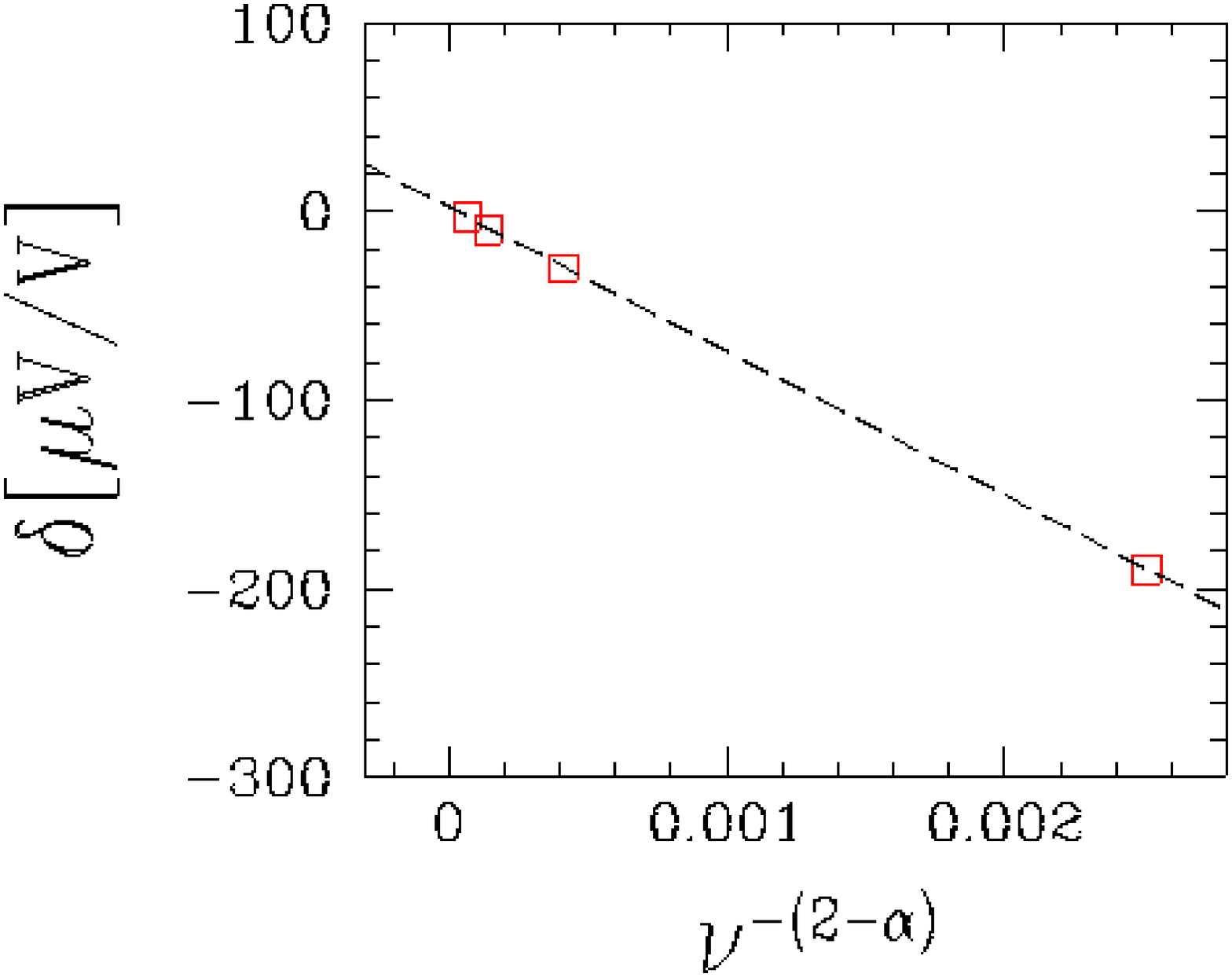}{D300mV}{{\it Linear fits of 792A data at fixed voltage.}
      The picture shows a least square fit line for 792A AC-DC differences $\delta$ 
      (in $\mu$V/V) at fixed 300 mV, and frequencies $10<\nu< 40$ Hz.
      Uncertainties are of order of 10 $\mu$V/V, and are not shown.}

      \umafigura{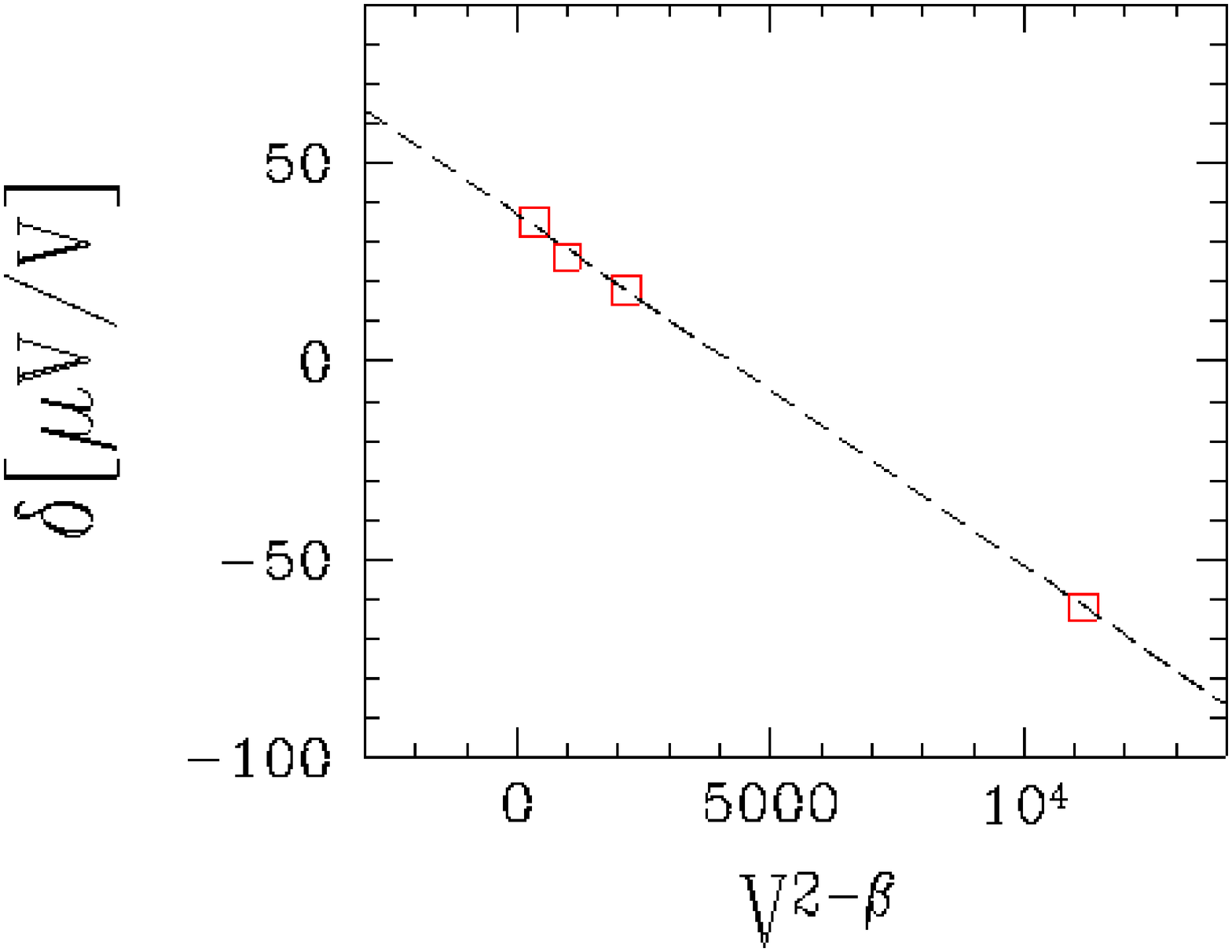}{D40Hz}{{\it Linear fits of 792A data at fixed frequency.}
      The picture shows a least square fit line for 792A AC-DC differences $\delta$ 
      (in $\mu$V/V) at fixed 40 Hz, and voltages $0.002< V< 0.020$ volts. 
      Uncertainties range from 35 to 75 $\mu$V/V, and are not shown.}

   Exponents  $\alpha$ and $\beta$ are adjusted under the condition of maximum 
   linear correlation between, for instance, fixed-voltage or fixed-frequency AC-DC 
   differences; this criterion has been consistently applied in similar adjustment
   on the context of scaling random systems \cite{fabio2001}.
   The implicit assumption behind Eq. \ref{SCALINGDELTA} is that the AC-DC difference
   is a smooth function of $Y$ and $\nu$, in the statistical sense.

      \savebox{1}[2in]{\vbox{\hsize=\linewidth\vsize=.5in\noindent\footnotesize
      This table shows the adjusted points $(\nu,\delta)$, their
      combined uncertainties $U$ and the interpolation errors $\Delta$
      for F792A calibration data.
      The frequency $\nu$ is given in hertz; $\delta$, $U$ and $\Delta$
      are expressed in $\mu V/V$.}}
      \begin{table}[!t]
        \caption{Frequency Interpolation of F792A Data at 0.300 V}
        \label{T300mV}
        \centering
        \begin{tabular}{ccc|c}
           $\nu$& $\delta$   &  U         &$\Delta$\\
                          \hline
           &&\\
           10   &   -190     &  10        & 0.03\\  
           20   &   -30      &  10        & 0\\
           30   &   -10      &  10        & -0.60\\  
           40   &   -3       &  10        & 0.60\\ 
           &&\\
        \end{tabular}\vskip 1em
        \begin{tabular}{c}
             \box1
        \end{tabular}
      \end{table}

      \savebox{1}[2in]{\vbox{\hsize=\linewidth\vsize=.5in\noindent\footnotesize
      This table shows the adjusted points $(V,\delta)$, their
      combined uncertainties $U$ and the interpolation errors $\Delta$
      for F792A calibration data.
      The voltage $V$ is given in volts; $\delta$, $U$ and $\Delta$
      are expressed in $\mu V/V$.}}
      \begin{table}[!t]
        \caption{Voltage Interpolation of F792A Data at 40 Hz}
        \label{T40Hz}
        \centering
        \begin{tabular}{ccc|c}
           $V$  & $\delta$   &  U         &$\Delta$\\
                                  \hline
           &&\\
           0.002&   -62      &  75        & -0.28\\  
           0.006&   18       &  60        & 0.27\\
           0.010&   26       &  50        & -1.87\\  
           0.020&   35       &  35        & 1.44\\ 
           &&\\
        \end{tabular}\vskip 1em
        \begin{tabular}{c}
             \box1
        \end{tabular}
     \end{table}

   Figs. \ref{D300mV} and \ref{D40Hz} show least square fits for AC-DC low-frequency differences
   on two different F792A instruments, obtained from calibration against PMTC, 
   both in voltage and frequency. Exponents $\alpha$ and $\beta$ were adjusted until 
   the correlation  coefficient $R^2$ of the fits came to a maximum. In Fig. \ref{D300mV}, 
   $\alpha=-0.6$ to $R^2=0.99998$; in Fig. \ref{D40Hz}, $\beta=3.5$ to $R^2=0.9995$ 
   (see Eq. \ref{SCALINGDELTA}). 

   The combined uncertainties $U$ are of 10 $\mu$V/V for the points in Fig. \ref{D300mV}; 
   for the points in Fig. \ref{D40Hz}, $U$ range from 35 to 75 $\mu$V/V. The standard errors of the
   fits are $\sigma=0.6$ $\mu$V/V (Fig. \ref{D300mV}) and $\sigma=1.7$ $\mu$V/V (Fig. \ref{D40Hz}).
   These figures represent negligible contribution to the combined uncertainties of  those points,
   and therefore no important contribution of the
   interpolation method to the uncertainty is expected in points interpolated
   between calibration points. The adjusted points $(\nu,\delta)$ and $(V,\delta)$, their
   combined uncertainties $U$ and the interpolation errors $\Delta$ 
   are summarized in Tables \ref{T300mV} and \ref{T40Hz}.

\section{Results for PMJTC Data}
\label{pmjtc}

   The AC standardization project at Inmetro has a fully operational automated PMJTC
   voltage system, and the first measurements are being carried out. This system is composed
   of two 180 $\Omega$-heater PMJTC, 
   calibrated at Physikalisch-Technische Bundesanstalt (PTB), 
   in 1.5 volts and frequencies from 10 to 1 MHz;
   twelve 90 $\Omega$, one 400 $\Omega$ and one 900 $\Omega$ PMJTC. These junctions
   are used with range-resistors from 200 mV to 1000 V; the standardization method
   chosen is the voltage step-up \cite{renato2009}.

      \umafigura{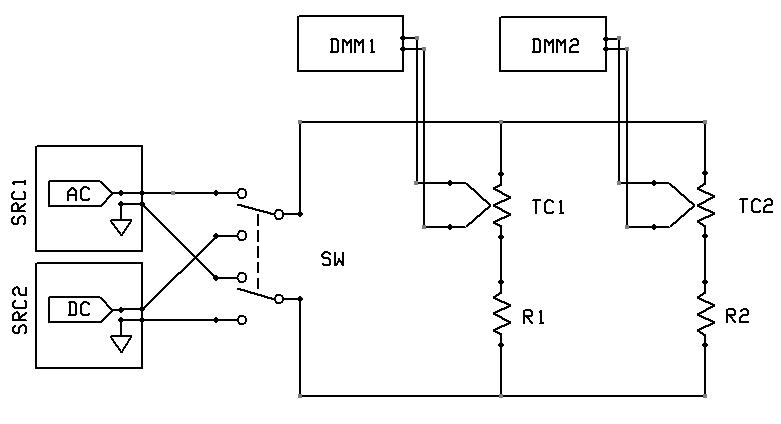}{PMJTC-I}{{\it PMJTC system at Inmetro.}
      This sketch shows the voltage system used to calibrate junctions in 50 V.
      {\tt SRC1} and {\tt SRC2} are multifunction calibrators;
      {\tt SW} is a switch; 
      {\tt DMM1} and {\tt DMM2} are digital nanovoltmeters;
      {\tt TC1/R1} and {\tt TC2/R2} are sets of PMJTC and range resistors
      (see text).}

   Fig. \ref{PMJTC-I} shows an schematics of the PMJTC voltage system used at Inmetro to
   calibrate junctions from 1.5 to 100 V. In this sketch,
   {\tt SRC1} and {\tt SRC2} are multifunction calibrators Fluke model 5720; 
   {\tt SW} is an AC-DC switching unit built at the Swiss Federal 
                          Office of Metrology and Accreditation (Metas);
   {\tt DMM1} and {\tt DMM2} are digital nanovoltmeters Keithley model 2182A;
   {\tt TC1} and {\tt TC2} are PMJTC made at PTB; and
   {\tt R1} and {\tt R2} are the proper range resistors for the set.

   We assume Eq. \ref{SCALINGDELTA} holds, and again adjust exponents $\alpha$
   and $\beta$ under the condition of maximum linear correlation between 
   AC-DC differences. Figs. \ref{MJ50V1} and \ref{MJ50V2} show frequency fits for 
   AC-DC low-frequency differences on two different 90 $\Omega$ PMJTC sets, obtained from 
   calibration against the 90 $\Omega$ standards for the 50 V step.
   Exponents $\alpha$ were adjusted until 
   the correlation  coefficient $R^2$ of the fits came to a maximum. In Fig. \ref{MJ50V1}, 
   $\alpha=-1.1$ to $R^2=0.893$; in Fig. \ref{MJ50V2}, $\alpha=0.5$ to $R^2=0.807$ 
   (see Eq. \ref{SCALINGDELTA}). 

      \umafigura{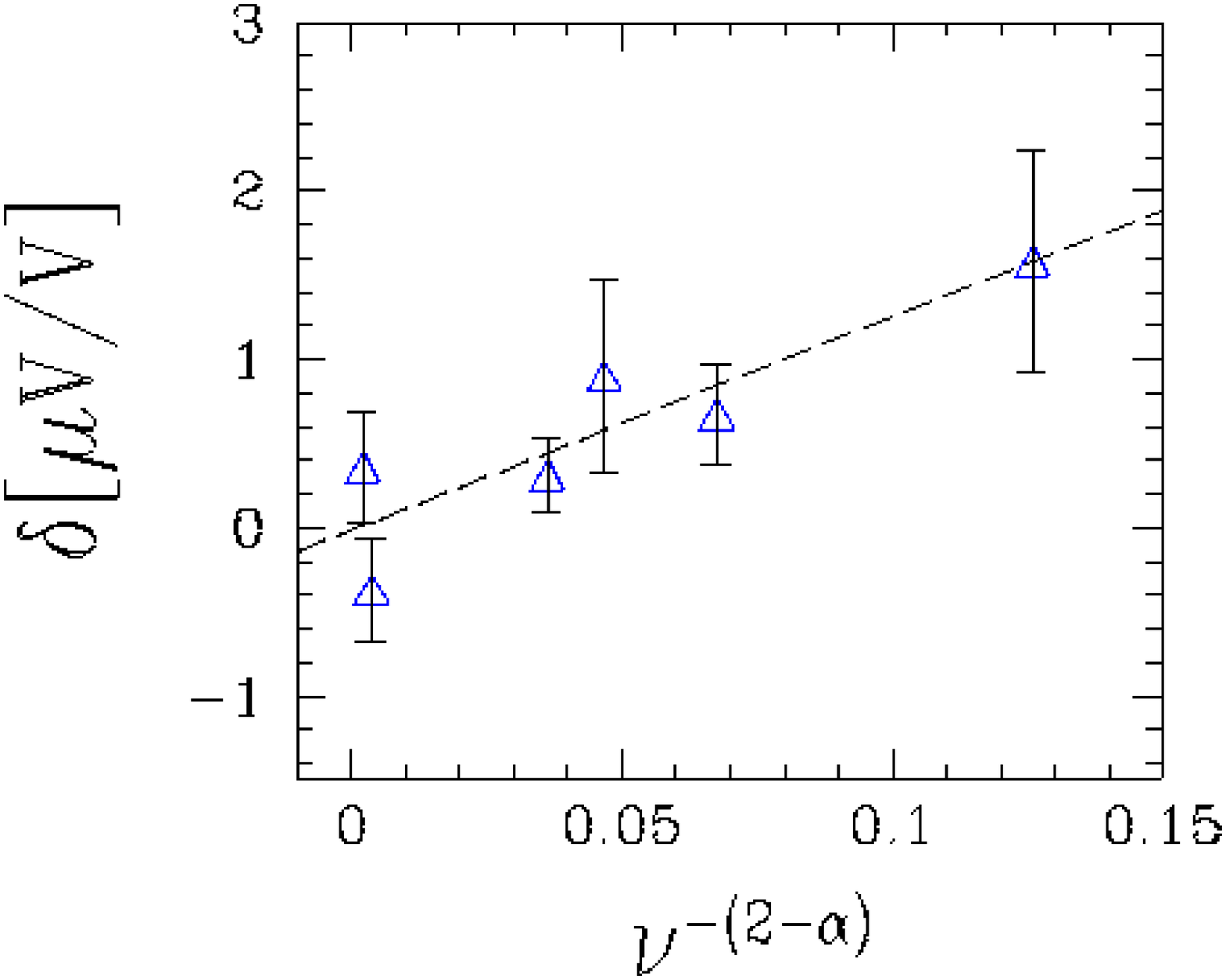}{MJ50V1}{{\it Linear fits of PMJTC data at fixed voltage (1).}
      The picture shows a least square fit of AC-DC differences $\delta$ (in $\mu$V/V)  
      for the 90 $\Omega$ PMJTC set {\tt 90-8/R100V}
      at fixed 50 V, and frequencies $10<\nu< 1000$ Hz. Error bars are standard deviations,
      and range from 0.2 to 0.7 $\mu$V/V.}

      \umafigura{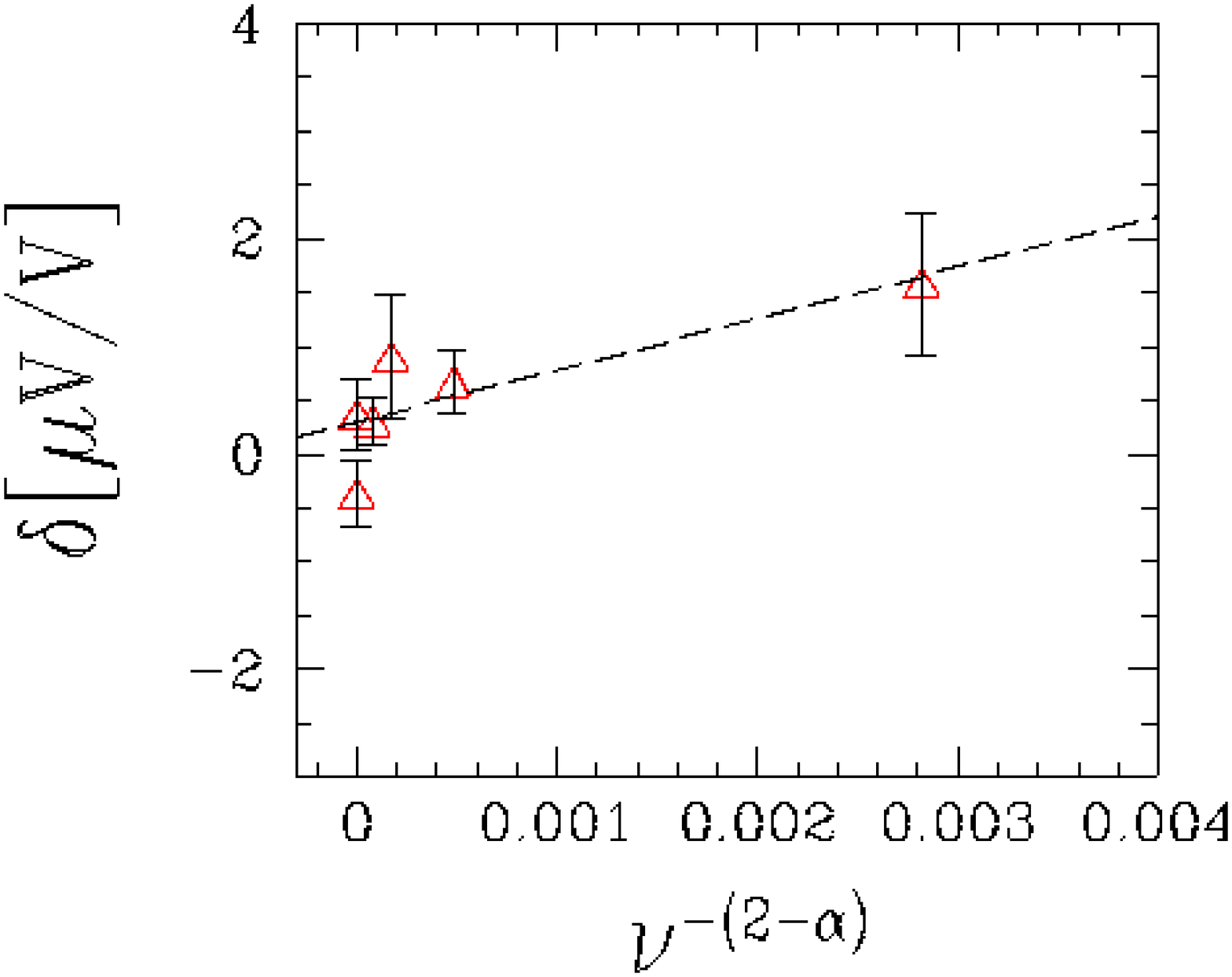}{MJ50V2}{{\it Linear fits of PMJTC data at fixed voltage (2).}
      The picture shows a least square fit of AC-DC differences $\delta$ (in $\mu$V/V)  
      for the 90 $\Omega$ PMJTC set {\tt 90-9/R300V}
      at fixed 50 V, and frequencies $10<\nu< 1000$ Hz. Error bars are standard deviations,
      and range from 0.4 to 0.7 $\mu$V/V.}

   The standard deviations $s$ of the AC-DC difference measurements range from 0.2 to 0.7 $\mu$V/V 
   for the points in Fig. \ref{MJ50V1};  for the points in Fig. \ref{MJ50V2}, $s$
   range from 0.4 to 0.7 $\mu$V/V. The standard errors of the
   fits are $\sigma=0.33$ $\mu$V/V (Fig. \ref{MJ50V1}) and $\sigma=0.15$ $\mu$V/V (Fig. \ref{MJ50V2}).
   These figures are of the order of the standard deviations of the measurements, and
   must represent no significant contribution to the combined uncertainties of those points,
   and therefore no important contribution of the
   interpolation method to the uncertainty is expected in points interpolated
   between calibration points. The adjusted points $(\nu,\delta)$, the standard deviations $s$ 
   and the interpolation errors $\Delta$ are summarized in Tables \ref{T50V1} and \ref{T50V2}.

      \savebox{1}[2in]{\vbox{\hsize=\linewidth\vsize=.5in\noindent\footnotesize
      This table shows the adjusted points $(\nu,\delta)$, their
      combined uncertainties $U$ and the interpolation errors $\Delta$
      for PMJTC set labelled {\tt 90-8/R100V}.
      The frequency $\nu$ is given in hertz; $\delta$, $U$ and $\Delta$
      are expressed in $\mu V/V$.}}
      \begin{table}[!t]
        \caption{Frequency Interpolation of PMJTC Data at 50 V (1)}
        \label{T50V1}
        \centering
        \begin{tabular}{ccc|c}
           $\nu$& $\delta$   &  s         &$\Delta$\\
                             \hline
           &&\\
           10   &   1.58     &  0.66      & 0.01\\  
           20   &   0.67     &  0.30      & -0.16\\
           30   &   0.90     &  0.57      & -0.33\\  
           40   &   0.31     &  0.22      & -0.13\\ 
           500  &   -0.37    &  0.31      & -0.40\\  
           1000 &   0.36     &  0.33      & 0.35\\ 
           &&\\
        \end{tabular}\vskip 1em
        \begin{tabular}{c}
             \box1
        \end{tabular}
     \end{table}

      \savebox{1}[2in]{\vbox{\hsize=\linewidth\vsize=.5in\noindent\footnotesize
      This table shows the adjusted points $(\nu,\delta)$, their
      combined uncertainties $U$ and the interpolation errors $\Delta$
      for PMJTC set labelled {\tt 90-9/R300V}.
      The frequency $\nu$ is given in hertz; $\delta$, $U$ and $\Delta$
      are expressed in $\mu V/V$.}}
      \begin{table}[!t]
        \caption{Frequency Interpolation of PMJTC Data at 50 V (2)}
        \label{T50V2}
        \centering
        \begin{tabular}{ccc|c}
           $\nu$  & $\delta$   &  s         &$\Delta$\\
                                  \hline
           &&\\
           10   &   3.11     &  0.46      & -0.01\\  
           20   &   -0.62    &  0.41      & 0.16\\
           30   &   -1.46    &  0.65      & -0.16\\  
           40   &   -1.60    &  0.50      & -0.16\\ 
           500  &   -1.53    &  0.49      & 0.05\\  
           1000 &   -1.46    &  0.55      & 0.12\\ 
           \hline &&\\
        \end{tabular}\vskip 1em
        \begin{tabular}{c}
             \box1
        \end{tabular}
     \end{table}

\section{Final Remarks}
\label{conclusion}

    Fully- and semi-automated systems based on electronic thermal standards can
    make it very practical for laboratories to handle demands for large
    numbers of calibration points in many different ranges. Recently reported automated
    systems have overall calibration time limited only by the charge duration
    of embedded batteries \cite{bob}, and though AC-DC transfer services can often
    last a few weeks long, electronic thermal standards are very likely to be around for a 
    long time yet. So, one of the many features of good interpolation methods
    is a possible reduction on the number of calibrated points within ranges of 
    electronic standards such as the F792A, both in voltage and frequency.

   The method presented here has never been applied to the study of measurement data
   from TC, though it can be very useful in the analysis of many theoretical models 
   of statistical physics \cite{fabio2001,eu2007}.  As in those models,
   further investigation of the correction exponents (Eq. \ref{SCALINGDELTA})
   for PMJTC data can bring deeper understanding of these non-linearities and
   might lead to important insights on these effects, instead of the other way 
   around.

   We also obtained small interpolation uncertainties on PMJTC standards.
   For PMJTC, a robust interpolation method, which in some way takes into 
   account the physics behind the measurement system, together with
   calibration schemes such as described in \cite{youlden1963}, can help to
   spare the extensive use of standards, while still promoting improvement
   in results.

   We should expect deviations from square and inverse-square laws that rule
   AC-DC differences at low frequencies in such standards to be related mainly  
   to Thomsom heating and non-linear 
   contributions to emissivity and thermal diffusivity in the heater
   \cite{hermach1952,inglis1992,sasaki1999,laiz1999,oldham1997}, as well as
   non-linear contributions from circuitry in electronic 
   standards, though the contribution of
   sistematic fluctuations of statistical nature should also be investigated.

\section*{Acknowledgment}

   The authors would like to thank R. T. B. Vasconcellos for helpful sugestions
   and a thorough revision of the manuscript.




\begin{references}





    \bibitem{laiz2005}
    Laiz,H., ``Low Frequency AC Voltage Standards: An Overview'', {\bf in}
    Proc. of VI SEMETRO, pp. 15 (2005)

    \bibitem{inglis1992}
    Inglis, B. D., ``Standards for AC-DC Transfer'', 
    {\it Metrologia} {\bf 29}, 191 (1992)


    \bibitem{calibration}
    Fluke Corp., {\it Calibration: Phylosophy in Practice}, Fluke Corp. (1994)

    \bibitem{cpem}
    F. A. Silveira, R. M. Souza. and R. P. Landim, ``Power-Law Picture for the 
    Interpolation of AC-DC Differences in Thermal Standards'', 
    Digest of Conf. on Precision Electromagnetic Measurements (CPEM2010), 
    June 13- June 18, 2010, Daejeon, Korea.

    \bibitem{hermach1952}
    Hermach, F. L., ``Thermal Converters as AC-DC Transfer Standards for Current
    and Voltage Measurements at Audio Frequencies'', 
    {\it J. Res. Nat. Bur. Std.} {\bf 48}, 163 (1952)


    \bibitem{sasaki1999}
    Sasaki, H. and Takahashi, K., ``Development of a High Precision AC-DC Transfer
    Standard using FRDC Method'',
    {\it Res. of Electr. Lab.} {\bf 989}, 90P (1999)

    \bibitem{apnote}
    Fluke Corp., {\it Design and Evaluation of the 792A AC/DC Transfer Standard}, 
    Fluke Application Note {\tt 1268781 A-ENG-N} Rev. B (2000)

    \bibitem{renato2009}
    Afonso Jr., R., Borghi, G., Landim, R. P., Afonso, E. and Silveira, F. A.,
    ``O Sistema de Padroniza\c c\~ao AC do Inmetro'', {\bf in}
    Proc. of VII Semetro, {\it N 52010} (2009)







   
    \bibitem{schoe1978}
    Schoenwetter, H. K., ``An RMS Digital Voltmeter/Calibratior for Very-Low 
    Frequencies'',
    {\it IEEE Trans. Instr. Meas.} {\bf 27}, 259 (1978)
   
    \bibitem{laiz1999}
    Laiz, M. L., ``Low Frequency Behaviour of Thin-Film Multijunction 
    Thermal Converters'', Physikalisch-Technische Bundesanstalt, PTB-Bericht 
    E-63 (1999)







   \bibitem{792apendixc}
   Fluke Corp., {\it Calculating Correction Factors for F$<$100 Hz},
   792A Transfer Standard Instruction Manual, PN 871723 Rev 1, C-1 (1990)
    

   \bibitem{fabio2001}
   Aar\~ao Reis, F. D. A., ``Universality and 
   Corrections to Scaling in the Ballistic Deposition Model'',
   {\it Phys. Rev. E} {\bf 63}, 056116 (2001)


   \bibitem{oldham1997}
   Oldham, N. M., Svetlana, A.-Z. and Parker, M. E., ``Exploring the Low-Frequency 
   Performance of Thermal Converters Using Circuit Models and a Digitally 
   Synthesized Source'', {\it IEEE Trans. Instr. Meas.} {\bf 46}, 352 (1997)

   \bibitem{eu2007}
   Silveira, F. A. and Aar\~ao Reis, F. D. A.,``Surface and Bulk Properties of 
   Deposits Grown With a Bidisperse Ballistic Deposition Model'',
   {\it Phys. Rev. E} {\bf 75}, 061608 (2007)





   \bibitem{bob}
   Teixeira, V. M., Ventura, R. V. F., Souza, R. M., Afonso Jr., R., Landim, R. P. and
   Afonso, E., ``Moving Towards Full Automation of High Accuracy Multifunction
   Instruments Calibration Systems'', {\bf in}
   Proc. of VIII Semetro, {\it N 52343} (2009)

   \bibitem{youlden1963}
   Youlden, W. J., ``Measurement Agreement Comparisons'', {\bf in}
   {\it Precision Measurement and Calibration: Statistical Concepts and Procedures},
   Ed. H. H. Ku, {\it NBS Special Publication} {\bf 300} vol {\bf 1}, 146 (1963)




\end{references}
\end{document}